\documentclass[aps,prl,showpacs,notitlepage,twocolumn,superscriptaddress,nofootinbib,preprintnumbers]{revtex4-2}
\usepackage{bbm}
\usepackage{mathrsfs}
\usepackage{epsfig}
\usepackage{soul,xcolor}
\usepackage{graphicx}
\usepackage{amsfonts}
\usepackage{amsthm}
\usepackage[figuresright]{rotating}
\usepackage{amssymb}
\usepackage{amsmath}
\usepackage{dcolumn}
\usepackage{physics}
\usepackage{float}
\usepackage{bm}
\usepackage{physics}
\usepackage{verbatim}
\usepackage{braket}
\usepackage[normalem]{ulem}
\usepackage[ruled,vlined,linesnumbered]{algorithm2e}
\usepackage{setspace}
\usepackage{lipsum}
\usepackage[colorlinks,linkcolor=blue,anchorcolor=blue,citecolor=blue,urlcolor=blue]{hyperref}
\renewcommand{\v}[1]{{\mathbf{\boldsymbol{#1}}}}

\newcommand{\br}{\bold{r}}
\newcommand{\psit}{\psi_{\theta}}

\begin{document}

\title{Simulating moir\'e quantum matter with neural network}

\author{Di~Luo}
\affiliation{Department of Physics, Massachusetts Institute of Technology, Cambridge, MA 02139, USA}
\affiliation{Department of Physics, Harvard University, Cambridge, MA 02138, USA}
\affiliation{The NSF AI Institute for Artificial Intelligence and Fundamental Interactions}
\author{David~D.~Dai}
\affiliation{Department of Physics, Massachusetts Institute of Technology, Cambridge, MA 02139, USA}
\author{Liang~Fu}
\affiliation{Department of Physics, Massachusetts Institute of Technology, Cambridge, MA 02139, USA}

\date{\today}

\begin{abstract}
Moiré materials provide an ideal platform for exploring quantum phases of matter. However, solving the many-electron problem in moir\'e systems is challenging due to strong correlation effects. We introduce a powerful variational representation of quantum states, many-body neural Bloch wavefunction, to solve many-electron problems in moiré materials accurately and efficiently. Applying our method to the semiconductor heterobilayer  WSe$_2$/WS$_2$,  we obtain a generalized Wigner crystal at filling factor $n=1/3$, a Mott insulator $n=1$, and a correlated insulator with local magnetic moments and antiferromagnetic spin correlation at $n=2$. Our neural network approach improves the simulation accuracy of strongly interacting moiré materials and paves the way for discovery of new quantum phases with variational learning principle in a unified framework.
\end{abstract}

\maketitle

\textit{Introduction---} The study of moiré materials is an exciting frontier in condensed matter physics \cite{moiremarvels}. These materials, which are artificial solids with a long-wavelength superlattice modulation, present a highly tunable platform for exploring novel quantum phases of matter \cite{BistritzerMoire}. The ability to tune electron density  and electronic miniband makes it possible to engineer novel quantum states with unprecedented control. In recent years, moir\'e heterostructures of semiconductor transition metal dichalcogenides (TMDs) \cite{Mak_Shan_2022} have proven to be a remarkable material platform hosting a plethora of correlated and topological electron phases, including Mott insulators \cite{mott1, mott2, mott3, mott4, mott5, mott6}, electron solids \cite{mott3, wigner1, wigner2, wigner3, wigner4, wigner5, wigner6, wigner7, wigner8, reddy2023artificial}, heavy Fermi liquids \cite{heavy1, heavy2, heavy3}, spin polaron metals \cite{pseudogap1, pseudogap2, pseudogap3}, integer and fractional quantum anomalous Hall states \cite{anomalous2, anomalous3, anomalous4, anomalous5, anomalous6, anomalous7, anomalous8, anomalous9, anomalous10, anomalous11}, fractional quantum spin Hall states \cite{spinhallevidence}, and unconventional superconductors \cite{guo2024superconductivity, xia2024unconventional}. 

Treating interacting electrons in moiré materials is challenging due to the presence of multiple energy scales and strong electron correlation effects \cite{reddy2023artificial}. While the continuum model approach has proven to be remarkably successful in describing noninteracting moir\'e band structure and band topology~\cite{mott1, wu2019topological}, 
numerical solution of the full continuum model Hamiltonians including electron-electron interaction is a difficult task~\cite{zhang2020density,yang2024metal}. The Hartree-Fock method fails to take into account electron correlation, while exact diagonalization studies are limited to small system size and very few moir\'e bands~\cite{morales2021metal,abouelkomsan2024band,yu2024fractional}. Meanwhile, traditional variational wavefunctions cannot capture distinct electron phases in a unified manner.         

Recent advancements in artificial intelligence and neural network quantum states have provided powerful new tools for tackling the quantum many-body problem \cite{Carleo602}. It has been shown that neural networks can efficiently represent a variety of quantum states, even those with high entanglement~\cite{Deng_2017, gao2017efficient, Glasser_2018, Levine_2019, sharir2021neural, luo_inf_nnqs}, with promising applications in simulating ground state~\cite{chen2023autoregressive,robledo2022fermionic, chen2022simulating, doi:10.1126/science.aag2302, Hibat_Allah_2020, PhysRevLett.124.020503, Irikura_2020, PhysRevResearch.3.023095, Han_2020,ferminet,Choo_2019,rnn_wavefunction,paulinet,Glasser_2018,Stokes_2020,Nomura_2017,martyn2022variational,Luo_2019,PhysRevLett.127.276402, https://doi.org/10.48550/arxiv.2101.07243,luo2022gauge}, finite temperature and real-time dynamics~\cite{xie2021ab,wang2021spacetime,py2021, gutierrez2020real, Schmitt_2020,Vicentini_2019,PhysRevB.99.214306,PhysRevLett.122.250502,PhysRevLett.122.250501,luo_gauge_inv,luo_povm} physics. To model fermionic quantum systems, which require antisymmetry, neural network backflow~\cite{Luo_2019} on Slater determinants and Bogoliubov-de Gennes wavefunctions was introduced to represent fermionic many-body wavefunctions. The idea was further developed by FermiNet~\cite{ferminet} and PauliNet~\cite{paulinet} for applications in quantum chemistry, demonstrating performance that surpasses conventional methods such as coupled cluster theory. Recent progress in continuum space simulations of condensed matter systems have also shown the potential of fermionic neural networks~\cite{cassella2023discovering,luo2023artificial,pescia2023message,kim2023neural,lou2023neural,entwistle2023electronic,wilson2022wave,li2022ab,scherbela2022solving,adams2021variational,smith2024ground,luo2023pairing}.

\begin{figure}[t!]
\vspace{6pt}
  \centering
  \includegraphics[width=\columnwidth]{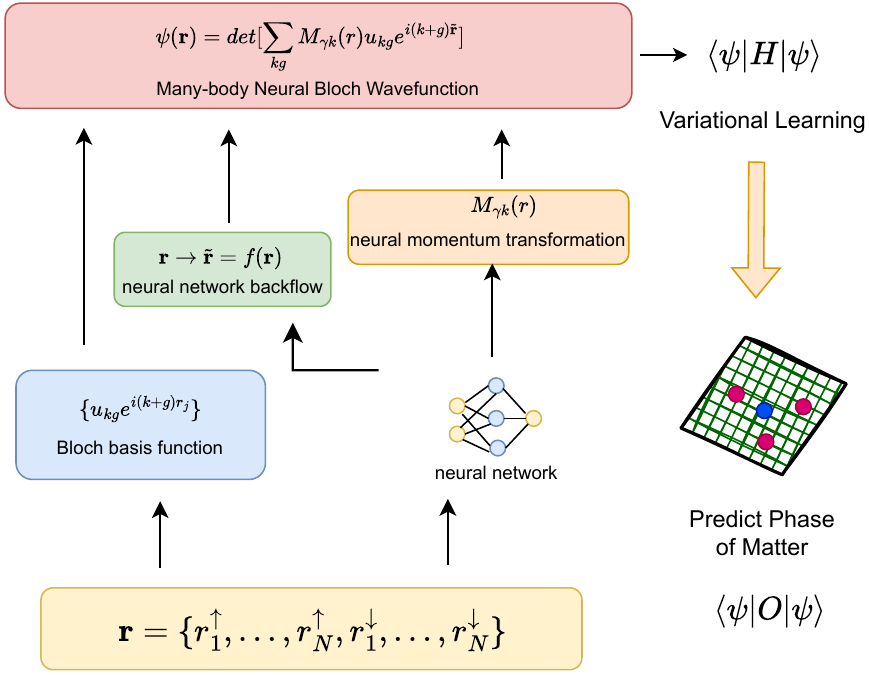}
  \caption{Many-body neural Bloch wavefunction for simulating moir\'e quantum materials.}
  \label{fig:nn}
\end{figure}

In this work, we introduce a new neural wavefunction for solving many-electron problems in moir\'e materials. Specifically, we study the  continuum model of TMD heterobilayers, which describes interacting electrons in a long-wavelength periodic potential. 
To capture the periodicity of the moir\'e lattice and the anti-symmetric property of fermionic wavefunction, 
we introduce the many-body neural Bloch wavefunction that generalizes the single-particle Bloch wavefunction by incorporating the many-body correlation with neural network backflow and neural momentum transformation via message passing graph network. To learn the many-body neural Bloch wavefunction, we adopt a first-principle variational approach with only input information from the moir\'e Hamiltonian and then compute different physical observables across various phases. We benchmark our approach over the Hartree Fock method and demonstrate its superior performance in ground state energy calculation. 

We perform neural wavefunction simulation of WSe$_2$/WS$_2$ at various fillings, accurately finding the ground-state energy and charge density distribution in the generalized Wigner crystal at $n=1/3$ and the charge-transfer Mott insulator at $n=1$, where $n$ is the filling factor defined as the number of holes per moir\'e unit cell. Furthermore,  we find that the system at $n=2$ hosts local magnetic moments and exhibits antiferromagnetic spin correlation.    
Our approach demonstrates the potential of simulating many-electron Hamiltonian and advancing our understanding of moir\'e quantum materials with a unified neural network framework.

\begin{figure}[t!]
    \centering
    \includegraphics[width=\columnwidth]{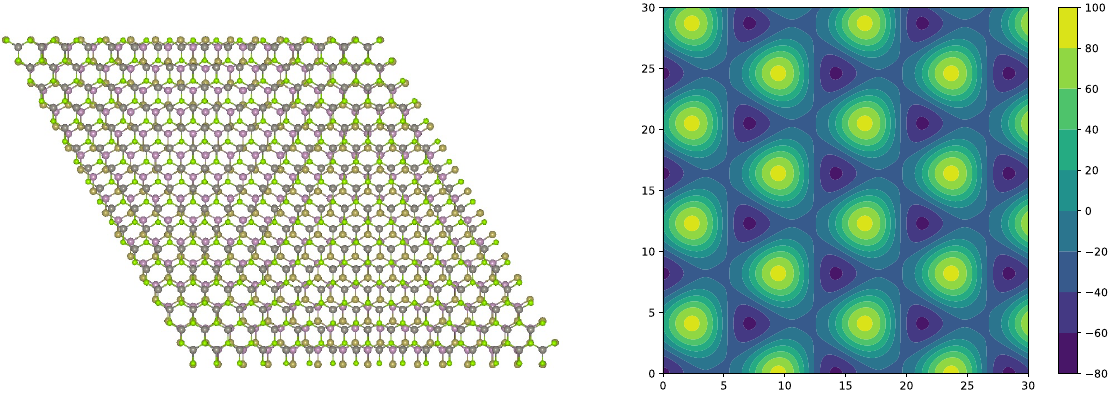}
    \caption{Depiction of semiconductor moiré superlattice (left). Color plots of the moiré potential with phase parameter $\phi=45^{\circ}$     equipotentials (right). } 
\label{fig:moire}
\end{figure}

\textit{Effective Continuum Hamiltonian---} We start with the continuum Hamiltonian for hole-doped TMD heterobilayers such as WSe$_2$/WS$_2$. Due to the large valence band offset between WSe$_2$ and WS$_2$ and the moir\'e superlattice from the 4$\%$ lattice mismatch, holes reside on the WSe$_2$ layer  and experience a long-wavelength periodic potential. The model Hamiltonian reads 
\begin{align}
&H = \sum_i \left( -\frac{\nabla_i^2}{2m} + V(r_i) \right) + \frac{1}{2} \sum_{i \ne j} \frac{e^2}{4\pi \epsilon_0 \epsilon |r_i - r_j|}, \label{eq:ham} \\
&V(r) = -2V_0 \sum_{l=1}^3 \cos (\mathbf{g}_l \cdot \mathbf{r} + \phi),
\end{align}
where $\mathbf{g}_l = \frac{4\pi}{\sqrt{3}a_M} (\cos \frac{2\pi l}{3}, \sin \frac{2\pi l}{3})$ are the reciprocal lattice vectors of the moiré superlattice. $\epsilon$ is the dielectric constant of the electrostatic environment. 
 For WSe$_2$/WS$_2$, the effective hole mass $m$ is approximately 0.5$m_e$. The dielectric constant of the electrostatic environment $\epsilon$ is estimated to be about 6 for hexagonal boron nitride substrate. The moire period $a_M$ varies with the twisted angle, and reaches 8.2 nm for aligned WSe$_2$/WS$_2$. 
 
Previous DFT study finds $V_0 = 15$ meV and $\phi = \pi/4$~\cite{mott2}. The phase parameter $\phi$ controls the moir\'e potential landscape. We have depicted the moir\'e potential of such system in Fig.~\ref{fig:moire}. It is important to note that the potential landscape shows one deep minimum and one shallow minimum at two high symmetry points in each moir\'e unit cell. 

As we will show by neural wavefunction simulation below, this interacting continuum Hamiltonian exhibits a rich phase diagram as a function of filling factor and twist angle encompasses a variety of electronic states, including charge-transfer Mott insulators and generalized Wigner crystals.

\textit{Many-body Neural Bloch Wavefunction---} We introduce a learnable many-body wavefunction for periodic potential potential system based on neural network backflow and neural momentum transformation on Bloch functions, which we call many-body neural Bloch wavefunction.

\begin{align}
    \Psi({\vb r}) &= det [\phi_\gamma (r_i; {\vb r})] \label{eq:det}\\
    \phi_\gamma(r_i; {\vb r}) &= \sum_{k,g} M_{\gamma k}(r_i; \vb r) u_{kg} e^{i(k+g) f(r_i;{\vb r})} \label{eq:blochnn}
\end{align}

The construction of Eq.~\ref{eq:blochnn} is illustrated in Fig.~\ref{fig:nn}. The many-body neural Bloch wavefunction generalizes the conventional Bloch function with many-body interaction via neural network backflow and neural momentum transformation. It includes three important ingredients: Bloch basis function, neural network backflow, and neural momentum transformation.

(i) Bloch basis function. A set of Bloch basis function $\{u_{kg} e^{i(k+g)r_j} \}$ is chosen based on the given moir\'e Hamiltonian, where $k$ is the mesh momentum and $g$ is the reciprocal lattice momentum, $u_{kg}$ is the Bloch function coefficient. The Bloch basis function encodes the basic physics information of the system and provides physics motivated structure for the construction of many-body neural Bloch wavefunction in the later steps. It guarantees that the many-body neural Bloch wavefunction is able to represent the Bloch wavefunction in the non-interacting limit or in the Hartree Fock method if $M$ and $f$ are identity function in Eq.~\ref{eq:blochnn}. In practice, it also enables pre-training by solving $u_{kg}$ with Hartree Fock or other methods to enhance the features in the Bloch basis function.

(ii) Neural network backflow. The idea of backflow is first proposed by Richard Feynman~\cite{feynman1954atomic} and further generalized to neural network backflow~\cite{Luo_2019}. Here we consider a neural network backflow $\vb r \rightarrow \vb {\tilde{r}} = f(\vb r)$ in position space, which will effectively introduce a backflow in the orbital space. Recently, message passing graph network have been used to provide neural network backflow in position space~\cite{pescia2023message,kim2023neural,luo2023pairing}. The message passing neural network backflow starts with a graph representation of the particle configurations as a graph $G^{0}=\{V^{0},E^{0}\}$ with node feature $V^0_i$ and edge feature $E^0_{ij}$. We initialize $V^0_i$ and $E^0_{ij}$ using the Fourier transformation features with moir\'e lattice and superlattice momentum for $\vb r_i$ and $\vb r_i - \vb r_j$ respectively (see Supplementary Materials for details). The message passing graph network starts with the initial graph features and iteratively update the graph for a certain number of times $l$ and reach a new set of graph features $V^{l}_i$ and $E^{l}_{ij}$. The physics motivation for message passing graph network is to capture the correlation among particles through iteration while preserve permutation symmetry. Notice that $V_i^l$ is a function of the many-body configuration $\vb r$. The neural network backflow is then given by:

\begin{equation}
    \vb{\tilde{r_i}} = \vb r_i + W V_i^l 
\end{equation}
where $W$ is complex-valued linear layer. The details of the message passing network structure used in this work is provided in the Supplementary Materials. 

\begin{figure}[t!]
  \centering
  \includegraphics[width=0.75\columnwidth]{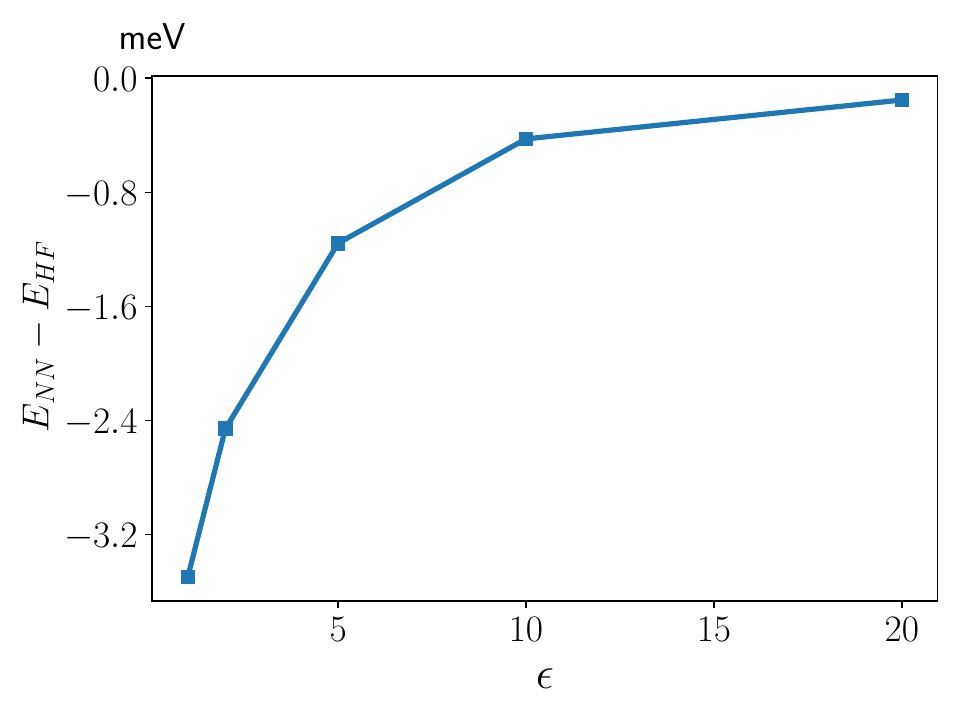}
  \caption{Energy per particle comparison between the many-body neural Bloch wavefunction and Hartree Fock of WSe$_2$/WS$_2$ with $V=15$ meV,  $a_M=8.2$ nm, $m^{*}=0.5 m_e$,  $\phi=45^\circ$ at filling $n=1$ with a $4 \times 4$ unit cell.}
  \label{fig:nnhf}
\end{figure} 

(iii) Neural momentum transformation. To further enhance the representation power of the many-body neural Bloch wavefunction, we introduce the neural momentum transformation as an important new feature:

\begin{equation}
    M_{\gamma k}(r_i; \vb r) = F_R([k_{\gamma}+g_{\gamma}, V^l_i]) * C_{\gamma k} 
\end{equation}
where $C_{\gamma k}$ is a linear layer, $F_R$ is a two layer fully connected neural network, $k_{\gamma}$ and $g_{\gamma}$ are the $k$ and $g$ points corresponding to the $\gamma$ label, * indicates the element-wise multiplication. The physics motivation of neural momentum transformation is to transform the existing momentum $k$ points into a new set of momentum $\gamma$ with many-body correlation, which captures the large momentum effect. The details of the transformation can be referred to the Supplementary Materials.

Our many-body neural Bloch wavefunction provides a general approach for simulating periodic quantum systems. For a given Hamiltonian of a periodic system, we can solve the corresponding Bloch basis function and build the many-body neural Bloch wavefunction following the above procedure. To include both spin up and spin down, we replace the Bloch basis function $\{u_{kg} e^{i(k+g)r_j} \}$ with $\{u_{k_{\sigma} g} e^{i(k_{\sigma}+g)r_j} \}$ of additional spin index $\sigma$.

To simulate the physics in the moir\'e system, we only utilize the information from the moir\'e Hamiltonian and learn the many-body neural Bloch wavefunction based on the variational principle. The learning of the many-body neural Bloch wavefunction $\psit(\br)$ with parameter $\theta$ is based on minimizing the energy according to Eq.~\ref{eq:energy}.

\begin{equation}
    E(\theta) = \langle \psit | H | \psit \rangle = \frac{\int  \psit^{*}(\br) H \psit(\br) d\br}{\int  \psit^{*}(\br) \psit(\br) d\br}. 
    \label{eq:energy}
\end{equation}

The above energy can be efficiently estimated using Monte Carlo sampling and the parameter update of $\theta$ is done with natural gradient method~\cite{sorella1998green}. Each single evaluation cost of the many-body neural Bloch wavefucntion in a $N$ particle system scales as $O(N^3)$ due to the complexity of determinant and the total optimization cost per step scale as $O(N^4)$ since the the Markov chain may take $O(N)$ evaluation of the many-body neural Bloch wavefucntion to equilibriate. After we obtain the many-body neural Bloch wavefunction $\psit(\br)$, we can compute any local observable efficiently from it using Monte Carlo sampling and investigate the corresponding physics. The polynomial scaling offers a feasible approach to simulate many-electron problems in the moir\'e system. The details of the optimization is provided in the Supplementary Materials.

\begin{figure*}[t!]
  \centering
  \includegraphics[width=2\columnwidth]{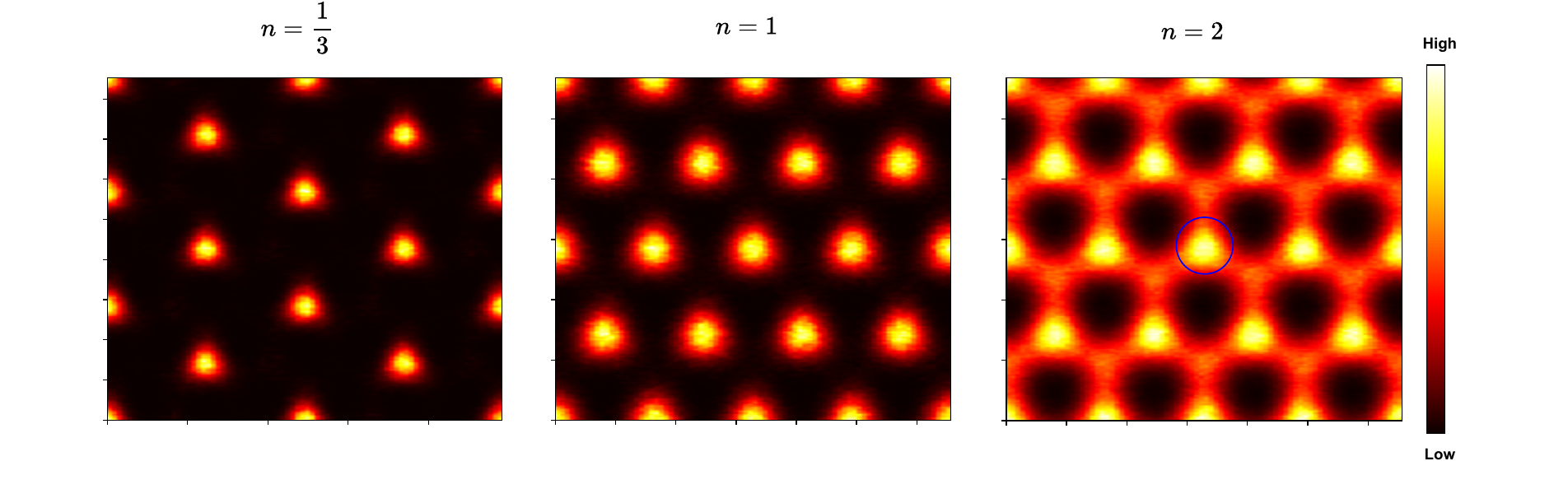}
  \caption{Charge density of WSe$_2$/WS$_2$ with $V=15$ meV,  $a_M=8.2$ nm, $m^{*}=0.5 m_e$,  $\phi=45^\circ$ at fillings $n=\frac{1}{3},1,2$.}
  \label{fig:density}
\end{figure*}

\textit{Results---} We first compare the performance of our many-body neural Bloch wavefunction with variational Hartree Fock calculation. We benchmark our results on $4 \times 4$ unit cells at filling $n=1$ with Coulomb interaction strength $1/\epsilon$ varied. The results are shown in Fig.~\ref{fig:nnhf}. It demonstrates from strong interaction $\epsilon=1$ to weak interaction $\epsilon=20$, the many-body neural Bloch wavefunction outperforms the Hartree Fock methods. The reduction in energy of the many-body neural Bloch wavefunction increases with the interaction strength. Clearly, the many-body neural Bloch wavefunction provides a more powerful tool for studying many-electron systems compared to conventional Hartree Fock method.

\begin{figure}[b!]
  \centering
  \includegraphics[width=\columnwidth]{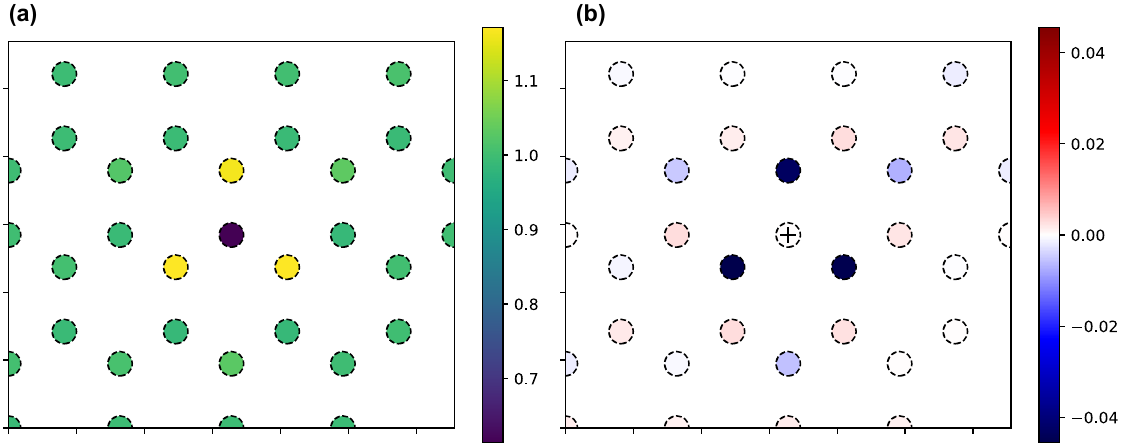}
  \caption{(a) spin up and spin down real-space pair correlation function (b) $S_{zz}$ two-point correlation function of WSe$_2$/WS$_2$ with $V=15$ meV,  $a_M=8.2$ nm, $m^{*}=0.5 m_e$,  $\phi=45^\circ$ at filling $n=2$. The color on each dot area $D_i$ in (a) and (b) represents the pair correlation value $g_{\uparrow \downarrow} (\v r \in D_i, \v r' \in D_o)$ and the two point correlation value $S_z(\v r \in D_i, \v r' \in D_o)$, where $D_o$ is the dot area centered in the origin at the middle of the figure.}
  \label{fig:gud}
\end{figure} 

Next, we apply the many-body neural Bloch wavefunction to study the ground states of WSe$_2$/WS$_2$ at various filling factors, using the model parameters $V=15$ meV,  $a_M=8.2$ nm, $m^{*}=0.5 m_e$,  $\phi=45^\circ$~\cite{zhang2020moire} and dielectric constant  $\epsilon=5$. We study various fillings $n=1/3, 1, 2$, and compute the charge density, which are shown in Fig.~\ref{fig:density}. Our simulations are performed in a $6 \times 6$ unit cell for $n=1/3$ and a $4 \times 4$ unit cell for $n=1,2$. 

At fractional filling $n=1/3$, the holes minimize mutual Coulomb interaction by occupying $1/3$ of the deep potential minima and forming a $\sqrt{3}\times \sqrt{3}$ structure. Thus, the ground state is a generalized Wigner Crystal phase as observed in the recent STM experiment~\cite{wigner4}. For filling $n=1$, the holes completely occupy the potential minima and 
form a Mott insulator. Since charges are strongly localized at $n=1/3$ and $1$, the exchange interaction between local moments is very small and the spins are fully polarized by a small magnetic field \cite{mott4}. Hence these simulations are performed for the fully spin polarized case.    
Our results from the many-body neural Bloch wavefunction accurately capture the strongly interacting regime from first principle based on only information in the Hamiltonian, opening up a new opportunity for quantitatively computing physical observables and comparing with experiments.

At $n=2$,  
we find the energy per particle with $S_z=0$ is more than $1.5$ meV lower than that of the fully polarized state, suggesting that the ground state is spin unpolarized. We further compute the ground state charge density, which is shown in Fig.~\ref{fig:density}. Interestingly, there exist both localized charges centered at the deep potential minima (A) and additional charges that spread around the secondary potential minima (B). If we take a circle centered at A with a diameter equal to the distance from A to the adjacent B, we find that the total charge enclosed is approximately 1. This finding suggests that every A site accommodates one particle and the other half of particles transfer to regions around B sites in order to avoid strong on-site Coulomb repulsion, consistent with the theoretical picture of charge transfer at $n>1$ \cite{mott2}. Therefore we conclude that due to strong Coulomb interaction, WSe$_2$/WS$_2$ at $n=2$ hosts local magnetic moments and therefore cannot be a simple band insulator.  

We further calculate the pair correlation function between opposite spins according to Eq.~\ref{eq:pair}, 
\begin{equation}\label{eq:pair}
    g_{\uparrow \downarrow}(\vb r,\vb r') = \frac{\int |\Psi(\v R)|^2 \sum_{i,j} \delta^{(2)}(\v r^{\uparrow}_i - \v r) \delta^{(2)}(\v r^{\downarrow}_j - \v r') \text{d}\v R}{n_{\uparrow}(\vb r) n_{\downarrow}(\vb r') \int |\Psi(\v R)|^2 \text{d}\v R},
\end{equation}
where $n_{\sigma}(\vb r)$ is the $\sigma$-particle density ($\sigma=\uparrow$ for spin up and $\sigma=\downarrow$ for spin down) given by the following: 
\begin{equation}\label{eq:pair}
    n_{\sigma}(\vb r) = \frac{\int |\Psi(\v R)|^2 \sum_{i} \delta^{(2)}(\v r^{\sigma}_i - \v r)  \text{d}\v R}{\int |\Psi(\v R)|^2 \text{d}\v R},
\end{equation}
We also compute the two-point spin correlation function.
\begin{equation}\label{eq:tpt}
    S_{zz}(\vb r,\vb r') = \frac{\int |\Psi(\v R)|^2 \sum M(\v r) M(\v r') \text{d}\v R}{ \int |\Psi(\v R)|^2 \text{d}\v R} - S_z(\v r)  S_z(\v r')
\end{equation}
where $S_z(\v r) =  (n_{\uparrow}(\vb r) - n_{\downarrow}(\vb r))/2$, $M(\v r) = (\sum_i \delta^{(2)}(\v r^{\uparrow}_i - \v r) - \sum_i \delta^{(2)}(\v r^{\downarrow}_i - \v r))/2$.

The results are shown in Fig.~\ref{fig:gud}. Each dot in the plot represents a circular area $D_i$  with diameter equal to the distance between adjacent potential minima (as shown by the circle in Fig.~\ref{fig:density}). The calculated pair correlation and spin correlation are given by $g_{\uparrow \downarrow} (\v r \in D_i, \v r' \in D_o)$ and $S_{zz}(\v r \in D_i, \v r' \in D_o)$, where we collect statistics for $\v r$ and $\v r'$ in the corresponding areas $D_i$ and $D_o$ (centered at the origin). 
The colors on each dot in Fig.~\ref{fig:gud} (a) and (b) represent the values of the two correlation functions ($D_i=D_o$ is excluded for the $S_{zz}$). It is clear that the spin up and spin down pair correlation reaches minimum at the origin and becomes maximum in adjacent potential minima. Also, the $S_{zz}$ plot shows strong antiferromagnetic spin correlation between adjacent A and B sites.    
We note that a recent optical study of WSe$_2$/WS$_2$ indeed observed local magnetic moments with antiferromagnetic exchange interaction at $n=2$ ~\cite{park2023dipole}, consistent with our simulation results.

\textit{Conclusion---} The study of moiré materials represents a groundbreaking frontier in quantum science, offering unprecedented opportunities to explore and manipulate novel quantum phases. Despite the significant challenges in simulating the many-electron problem within these complex systems, advancements in artificial intelligence and neural network-based wavefunctions have provided powerful new tools that surpass traditional methods in accuracy and flexibility. The development of the many-body neural Bloch wavefunction enhances our ability to model strongly interacting quantum states under periodic potential, thereby deepening our understanding of moiré materials. It holds the promise of uncovering new quantum phenomena and driving innovations in quantum computing and material science, solidifying the importance of moiré materials in future research and applications.

\textit{Note.} Near the completion of this work, we were informed of a related work~\cite{li2024emergent}. Here we develop a different architecture and method by introducing the many-body neural Bloch wavefunction that generalizes Bloch function via neural network backflow and neural momentum transformation. 

\textit{Acknowledgement.} The authors appreciate helpful discussions with Timothy Zaklama, Aidan Reddy, Ahmed Abouelkomsan, Ryan Levy, and Zhihuan Dong. We acknowledge the MIT SuperCloud for providing the computing resources used in this paper. 
This work is primarily supported by Simons Investigator Award from the Simons Foundation. DDD was supported by the Undergraduate Research Opportunities Program at MIT. DL acknowledges support from the NSF AI Institute for Artificial Intelligence and Fundamental Interactions (IAIFI) and the U.S. Department of Energy, Office of Science, National Quantum Information Science Research Centers, Co-design Center for Quantum Advantage (C2QA) under contract number DE-SC0012704. 

\newpage

\bibliography{references}
\bibliographystyle{apsrev4-1}

\appendix

\clearpage

\onecolumngrid
\begin{center}
	\noindent\textbf{Supplementary Material}
	\bigskip
		
	\noindent\textbf{\large{}}
\end{center}

\section{I. Details on Neural Network Architecture}

A message passing graph network starts with a graph representation of the particle configurations as a graph $G^{0}=\{V^{0},E^{0}\}$ with node feature $V^0_i$ and edge feature $E^0_{ij}$. In this work, we build the message passing graph network based on the spirit in Ref.~\cite{pescia2023message} with a few features added. We input the particle configuration into a graph by defining a set of vertex vectors $V_i^{0}$ and edge vectors $E_{ij}^{0}$:
\begin{equation}
    V^{0}_i = (v^{0}_i(\v r_i, s_i),  h^0_i) \quad, \quad  E^{0}_{ij} = (e^{0}_{ij}(\v r_{ij}, s_i),  h^0_{ij}) 
\end{equation}
where each vertex and edge correspond to a particle and a pair of particles with indices $i$ and $j$ running from $1$ to $N$. $\v r_i$ is the $i$-th particle position, $\v r_{ij} = \v r_{i} - \v r_{j} $ is the displacement between particles $i$ and $j$, and $s_i=\pm 1$ for spin up and spin down respectively. $v^0_i (\v r_i, s_i)$ and $e^0_{ij}(\v r_{ij}, s_i)$ use a similar sin and cos embedding as Ref.~\cite{pescia2023message}, $h^0_i$ and $h^0_{ij}$ are randomly initialized parameters such that $V^0_i$ and $E^0_{ij}$ have feature dimensions both equal to 32.

The vertex and edge vectors are processed iteratively to generate $V^l_i$ and $E^l_{ij}$. The $l$-th iteration ($l \geq 2$) is:
\begin{align}
    Q_{ij} &= W_Q^l E_{ij}^{l-1}, \quad K_{ij} = W_K^l E_{ij}^{l-1},\\
    \label{Eq:vertex}
    m_{ij}^{l} &= F^{l}(\sum_n  Q_{in}  K_{nj}) \odot G^{l}( E_{ij}^{l-1}),\\
    h_{i}^{l} &= U^{l}([\sum_j m_{ij}^{l}, V_{i}^{l-1}]) + h_{i}^{l-1}, \\
    h_{ij}^{l} &= H^{l}([ m_{ij}^l,  E_{ij}^{l-1}]) + h_{ij}^{l-1}, \\
    V^{l}_i &= (v^{0}_i,  h^l_i) \quad, \quad E^{l}_{ij} = (e^{0}_{ij},  h^l_{ij})
\end{align}
where $\otimes$ is the tensor product, $\odot$ is the element-wise multiplication, $F^{l},G^{l},H^{l}, U^{l}$ are one-layer fully-connected neural networks with hidden dimension 32 and GELU activation function. $W^l_Q$ and $W^l_K$ are embedding matrices and we use $l=2$ iteration.

\section{II. Details on Wavefunction Construction and Optimization}

We further construct the many-body neural Bloch wavefunction based on the neural network output. Even though we use the message passing network in this work, the following construction is general for generic permutation invariant function $V_i^l$. To construct neural network backflow,
\begin{equation}
    \vb{\tilde{r_i}} = \vb r_i + W V_i^l 
\end{equation}

we utilize that $W$ is a complex-valued linear layer with output dimension 2 and no activation function. 

To construct neural momentum transform, 

\begin{equation}
    M_{\gamma k}(r_i; \vb r) = F_R([k_{\gamma}+g_{\gamma}, V^l_i]) * C_{\gamma k} 
\end{equation}

we utilize that $C_{\gamma k}$ is a linear layer of size $\gamma \times k$ with no activation function, $F_R$ is a two layer fully connected neural network with GELU activation function followed by multiplying a simple factor function taking the sum of all distance of the particles as input. $k_{\gamma}$ and $g_{\gamma}$ are the $k$ and $g$ points corresponding to the $\gamma$ label, * indicates the element-wise multiplication. In practice, we choose $\gamma$ to be the number of particles and $k$ to be the number of mesh points in the first band. 

To optimize the neural network, we use natural gradient method~\cite{sorella1998green}  which takes 1000 steps with learning rate $\text{lr}=10^{-3}$ for fully polarized case and $\text{lr}=3 \times 10^{-4}$ for spin unpolarized case. The number of MCMC samples during optimization is 4128. The optimization is implemented in JAX~\cite{jax2018github} with neural network initialization using LecunNormal~\cite{klambauer2017self}.

\end{document}